\documentclass{aa}

\usepackage[varg]{txfonts}
\usepackage{epsfig,graphicx,natbib,url,twoopt}
\usepackage[varg]{txfonts}
\usepackage{hyperref}          
\usepackage{pdfcomment,acronym}      
\hypersetup{
 colorlinks=true,  
 urlcolor=blue,    
 linkcolor=red     
}

\def\# #1\par{\par\mbox{}\\ \noindent{\color{red}\small $\sharp$ #1}\\} 

\bibpunct{(}{)}{;}{a}{}{,}    
\makeatletter
 \newcommandtwoopt{\citeads}[3][][]{%
   \nonstopmode
   \href{http://adsabs.harvard.edu/abs/#3}%
        {\def\hyper@linkstart##1##2{}%
         \let\hyper@linkend\@empty\citealp[#1][#2]{#3}}
   \biblink{#3}{\href{http://adsabs.harvard.edu/abs/#3}{ADS}}%
   \errorstopmode}            
 \newcommandtwoopt{\citepads}[3][][]{%
   \nonstopmode
   \href{http://adsabs.harvard.edu/abs/#3}%
        {\def\hyper@linkstart##1##2{}%
         \let\hyper@linkend\@empty\citep[#1][#2]{#3}}
   \biblink{#3}{\href{http://adsabs.harvard.edu/abs/#3}{ADS}}%
   \errorstopmode}            
 \newcommandtwoopt{\citetads}[3][][]{%
   \nonstopmode
   \href{http://adsabs.harvard.edu/abs/#3}%
        {\def\hyper@linkstart##1##2{}%
         \let\hyper@linkend\@empty\citet[#1][#2]{#3}}
   \biblink{#3}{\href{http://adsabs.harvard.edu/abs/#3}{ADS}}%
   \errorstopmode}            
 \newcommandtwoopt{\citeyearads}[3][][]{%
   \nonstopmode
   \href{http://adsabs.harvard.edu/abs/#3}%
        {\def\hyper@linkstart##1##2{}%
         \let\hyper@linkend\@empty\citeyear[#1][#2]{#3}}
   \biblink{#3}{\href{http://adsabs.harvard.edu/abs/#3}{ADS}}%
   \errorstopmode}            
\makeatother

\newcount\longrefs
\def\aap{\ifnum\longrefs=1 {Astron.\ Astrophys.}\else 
                           {A\hbox{\rm \&}A}\fi}
\def\aapr{\ifnum\longrefs=1 {Astron.\ Astrophys.\ Rev.}\else 
                            {A\hbox{\rm \&}AR}\fi}
\def\aaps{\ifnum\longrefs=1 {Astron.\ Astrophys.\ Suppl.}\else 
                            {A\hbox{\rm \&}A Suppl.}\fi}
\def\actaa{\ifnum\longrefs=1 {Acta Astronomica}\else
                            {Acta Astron.}\fi}
\def\aipcs{\ifnum\longrefs=1 {Am.\ Inst.\ Phys.\ Conf.\ Series}\else
                             {AIP Conf.\ Ser.}\fi}
\def\aj{\ifnum\longrefs=1 {Astron.\ J.}\else 
                          {AJ}\fi} 
\def\ao{\ifnum\longrefs=1 {Applied Optics}\else 
                           {Appl.\ Opt.}\fi} 
\def\aspcs{\ifnum\longrefs=1 {Astron.\ Soc.\ Pacific Conf.\ Series}\else 
                           {ASP Conf.\ Ser.}\fi} 
\def\apj{\ifnum\longrefs=1 {Astrophys.\ J.}\else 
                           {ApJ}\fi} 
\def\apjl{\ifnum\longrefs=1 {Astrophys.\ J. Lett.}\else 
                            {ApJL}\fi} 
\def\aplett{\ifnum\longrefs=1 {Astrophys.\ J. Lett.}\else 
                            {ApJ}\fi} 
\def\apjs{\ifnum\longrefs=1 {Astrophys.\ J. Suppl.}\else 
                            {ApJS}\fi}
\def\apss{\ifnum\longrefs=1 {Astrophys.\ and Space Science}\else 
                            {Astrophys.\ Space Sci.}\fi}
\def\araa{\ifnum\longrefs=1 {Ann.\ Rev.\ Astron.\ Astrophys.}\else 
                            {ARA\hbox{\rm \&}A}\fi}
\def\azh{\ifnum\longrefs=1 {Astronomicheskii Zhurnal}\else 
                            {Astron.\ Zhur.}\fi}
\def\baas{\ifnum\longrefs=1 {Bull.\ Am.\ Astron.\ Soc.}\else 
                            {BAAS}\fi}
\def\bain{\ifnum\longrefs=1 {Bull.\ Astronom.\ Institutes Netherlands}\else
                            {Bull.\ Astr.\ Inst.\ Neth.}\fi}
\def\cjaa{\ifnum\longrefs=1 {Chinese Jour.\ Astron.\ Astrophys.}\else 
                            {Chin.\ J.\ A\&A}\fi}
\def\gca{\ifnum\longrefs=1 {Geochim.\ Cosmochim.\ Acta}\else 
                           {Geochim.\ Cosmochim.\ Acta}\fi}
\def\grl{\ifnum\longrefs=1 {Geophys.\ Res.\ Lett.}\else 
                           {Geoph.\ Res.\ Lett.}\fi}
\def\iaucirc{\ifnum\longrefs=1 {IAU Circulars}\else 
                          {IAU Circ.}\fi}
\def\icarus{\ifnum\longrefs=1 {Icarus}\else 
                          {Icarus}\fi}
\def\ip{\ifnum\longrefs=1 {in press}\else 
                          {in press}\fi}
\def\jcap{\ifnum\longrefs=1 {Jour.\ Cosmology Astropart.\ Phys.}\else 
                          {JCAP}\fi}
\def\jgr{\ifnum\longrefs=1 {J.\ Geophys.\ Res.}\else 
                           {J.\ Geophys.\ Res.}\fi}  
\def\jrasc{\ifnum\longrefs=1 {J.\ Royal Astron.\ Soc.\ Canada}\else 
                           {JRAS Can.}\fi}  
\def\memsai{\ifnum\longrefs=1 {Mem.~Soc.~Astron.~Italiana}\else
                              {MmSAI}\fi}
\def\mnras{\ifnum\longrefs=1 {Mon.\ Not.\ Roy.\ Astron.\ Soc.}\else 
                             {MNRAS}\fi} 
\def\na{\ifnum\longrefs=1 {New Astronomy}\else 
                           {New Astron.}\fi}
\def\nar{\ifnum\longrefs=1 {New Astronomy rev.}\else 
                           {New Astron.\ Rev.}\fi}
\def\nat{\ifnum\longrefs=1 {Nature}\else 
                           {Nat}\fi}
\def\pasa{\ifnum\longrefs=1 {Pub.\ Astron.\ Soc.\ Australia}\else 
                            {PASA}\fi} 
\def\pasj{\ifnum\longrefs=1 {Pub.\ Astron.\ Soc.\ Japan}\else 
                            {PASJ}\fi} 
\def\pasp{\ifnum\longrefs=1 {Pub.\ Astron.\ Soc.\ Pacific}\else 
                            {PASP}\fi} 
\def\physscr{\ifnum\longrefs=1 {Physica Scripta}\else 
                            {Phys.\ Scrip.}\fi} 
\def\planss{\ifnum\longrefs=1 {Planetary \& Space Science}\else 
                            {Plan. \& Space Sci.}\fi} 
\def\procspie{\ifnum\longrefs=1 {Proc.\ SPIE}\else 
                            {Proc.\ SPIE}\fi} 
\def\qjras{\ifnum\longrefs=1 {Quarterly J.\ Royal Astron.\ Soc.}\else 
                            {QJRAS}\fi} 
\def\rmxaa{\ifnum\longrefs=1 {Revista Mexicana de Astron.\ y Astrofys.}\else 
                            {RMxAA}\fi} 
\def\sa{\ifnum\longrefs=1 {Soviet Astron..}\else 
                               {Sov.\ Astron.}\fi}
\def\skytel{\ifnum\longrefs=1 {Sky \& Telescope}\else 
                            {Sky \& Tel.}\fi} 
\def\solphys{\ifnum\longrefs=1 {Solar Phys.}\else 
                               {SoPh}\fi}
\def\sovast{\ifnum\longrefs=1 {Soviet Astronomy}\else 
                               {Sov.\ Ast.}\fi}
\def\ssr{\ifnum\longrefs=1 {Space Science Rev.}\else 
                               {Space\ Sci.\ Rev.}\fi}
\def\zap{\ifnum\longrefs=1 {Zeitschr.\ f.\ Astrophysik}\else
                               {Z.\ Astrophys.}\fi}

\makeatletter
\newcommand{\bibnote}[2]{\@namedef{#1note}{#2}}
\newcommand{\biblink}[2]{\@namedef{#1link}{#2}}
\makeatother

\newacro{AA}{Astronomy \& Astrophysics}
\newacro{ADS}{Astrophysics Data System}
\newacro{AIA}{Atmospheric Imaging Assembly}
\newacro{AO}{adaptive optics}
\newacro{ApJ}{Astrophysical Journal}
\newacro{AR}{active region}
\newacro{BFI}{Broad-band Filter Imager}
\newacro{CE}{coronal equilibrium}
\newacro{CfA}{Center for Astrophysics}
\newacro{CME}{coronal mass ejection}
\newacro{CRD}{complete redistribution}
\newacro{CRISP}{CRisp Imaging SpectroPolarimeter}
\newacro{CRISPEX}{CRisp SPectral EXplorer}
\newacro{CS}{coherent scattering}
\newacro{DEM}{Differential Emission Measure}
\newacro{DF}{dynamic fibril}
\newacro{DKIST}{Daniel K. Inouye Solar Telescope}
\newacro{DLR}{Deutsches Zentrum f\"ur Luft- und Raumfahrt}
\newacro{DOT}{Dutch Open Telescope}
\newacro{DST}{Richard B. Dunn Solar Telescope}   
\newacro{EB}{Ellerman bomb}
\newacro{EDP}{\'{E}dition Diffusion Presse Sciences}  
\newacro{EIT}{Extreme ultraviolet Imaging Telescope}
\newacro{EPIC}{European participation in Solar-C}
\newacro{ERC}{European Research Council}
\newacro{ESA}{European Space Agency}
\newacro{EST}{European Solar Telescope}
\newacro{EUV}{extreme ultraviolet}
\newacro{FAF}{flaring active-region fibril}
\newacro{FITS}{Flexible Image Transport System}
\newacro{FOV}{field of view}
\newacro{fov}{field of view}
\newacro{FWHM}{full width at half maximum}
\newacro{HAO}{High Altitude Observatory}
\newacro{HD}{hydrodynamics}
\newacro{Hi-C}{High Resolution Coronal Imager Sounding Rocket}
\newacro{HMI}{Helioseismic and Magnetic Imager}
\newacro{IAA}{Instituto de Astrof\'{i}sica de Andaluc\'{i}a}
\newacro{IAC}{Instituto de Astrof\'{i}sica de Canarias}
\newacro{IAS}{Institut d'Astrophysique Spatiale}
\newacro{IDL}{Interactive Data Language}
\newacro{IMaX}{Imaging Magnetograph eXperiment}
\newacro{INAF}{Istituto Nazionale di Astrofisica}
\newacro{IB}{IRIS bomb}
\newacro{IR}{infrared}
\newacro{IRIS}{Interface Region Imaging Spectrograph}
\newacro{ISAS}{Institute of Space and Astronautical Science}
\newacro{ISP}{Institute for Solar Physics}
\newacro{ISS}{International Space Station}
\newacro{ISSI}{International Space Science Institute}
\newacro{ITA}{Institute for Theoretical Astrophysics}
\newacro{JAXA}{Japan Aerospace Exploration Agency}
\newacro{KIS}{Kiepenheuer--Institut f\"{u}r Sonnenphysik}
\newacro{KPNO}{Kitt Peak National Observatory}
\newacro{LASP}{Laboratory for Atmospheric and Space Physics}
\newacro{LC}{liquid cristal}
\newacro{LMSAL}{Lockheed Martin Solar and Astrophysics Labratory}
\newacro{LOS}{line of sight}
\newacro{LTE}{local thermodynamic equilibrium}
\newacro{MC}{magnetic concentration}
\newacro{MCAO}{multi-conjugate adaptive optics} 
\newacro{MDI}{Michelson Doppler Imager}
\newacro{ME}{Milne-Eddington} 
\newacro{MHD}{magnetohydrodynamics}
\newacro{MOMFBD}{Multi-Object Multi-Frame Blind Deconvolution}
\newacro{MPE}{Max--Planck--Institut f\"ur extraterrestrische Physik}
\newacro{MPG}{Max--Planck--Gesellschaft}
\newacro{MPS}{Max Planck Institute for Solar System Research}
\newacro{MSSL}{Mullard Space Science Laboratory}
\newacro{MTF}{modulation transfer function}
\newacro{NAOJ}{National Astronomical Observatory of Japan}
\newacro{NASA}{National Aeronautics and Space Administration}
\newacro{NLTE}{non-local thermodynamic equilibrium}
\newacro{NLFFF}{non-linear force-free field}
\newacro{NOAA}{National Oceanic and Atmospheric Administration}
\newacro{non-E}{non-equilibrium}
\newacro{NSO}{National Solar Observatory}
\newacro{NWO}{Netherlands Organisation for Scientific Research}
\newacro{PRD}{partial redistribution}
\newacro{PROBA2}{PRoject for OnBoard Autonomy}
\newacro{PSF}{point spread function}
\newacro{QS}{quiet Sun}
\newacro{QSEB}{quiet-Sun Ellerman-like brightening} 
\newacro{RAL}{Rutherford Appleton Laboratory}
\newacro{RBE}{rapid blue-shifted excursion}
\newacro{R-MHD}{radiation hydrodynamics}
\newacro{rms}{root mean square}
\newacro{RMS}{root mean square}
\newacro{ROB}{Royal Observatory of Belgium}
\newacro{ROI}{region of interest}
\newacro{RRE}{rapid red-shifted excursion}
\newacro{RTE}{radiative transfer equation}
\newacro{SE}{statistical equilibrium}
\newacro{SB}{Saha Boltzmann}
\newacro{SDO}{Solar Dynamics Observatory}
\newacro{SJI}{slit-jaw image}
\newacro{SNR}{signal-to-noise ratio}
\newacro{SO}{Solar Orbiter}
\newacro{SoHO}{Solar and Heliospheric Observatory}
\newacro{SP}{Spectropolarimeter}
\newacro{SST}{Swedish 1-m Solar Telescope}
\newacro{SUMER}{Solar Ultraviolet Measurements of Emitted Radiation}
\newacro{SUFI}{Sunrise Filter Imager}
\newacro{SVD}{singular value decomposition}
\newacro{SVST}{Swedish Vacuum Solar Telescope}
\newacro{THEMIS}{T\'{e}lescope H\'{e}liographique pour l'Etude du 
   Magn\'{e}tisme et des Instabilit\'{e} Solaires}     
\newacro{TR}{transition region}
\newacro{TRACE}{Transition Region and Coronal Explorer}
\newacro{TSI}{total solar irradiance}
\newacro{UT}{Universal Time}
\newacro{UV}{ultraviolet}
\newacro{VAULT}{Very high angular resolution ultraviolet telescope}
\newacro{VIRGO}{Variability of solar IRradiance and Gravity Oscillations}
\newacro{VTT}{Vacuum Tower Telescope}    
\newacro{XRT}{X-Ray Telescope}

\def\acp#1{\pdftooltip{\acs{#1}}{\acl{#1}}}  

\long\def\startignore #1\stopignore{}   

\def\FeI{\mbox{Fe\,\specchar{i}}}

\def\Hmin{\hbox{\rmH$^{^{_-}}\!$}}      

\def\Halpha{\mbox{H\hspace{0.1ex}$\alpha$}} 

\def\CaIR{\mbox{Ca\,\specchar{ii}\,8542\,\AA}} 

\def\level #1 #2#3#4{$#1 \; ^{#2} \mbox{#3} ^{#4}$}   

\def\arcsec{\hbox{$^{\prime\prime}$}}

\def\kms{\hbox{km$\;$s$^{-1}$}}

\def\={\hbox{$\!=\!$}}                     

\def\kms{\hbox{km$\;$s$^{-1}$}}

\def\Halpha{\mbox{H\hspace{0.1ex}$\alpha$}} 

\def\FeI{\ion{Fe}{i}}
\def\FeIline{\ion{Fe}{i}~6173\,\AA}
\def\CaIR{\ion{Ca}{ii}~8542\,\AA}
\def\Hmin{\hbox{\rm H$^{^{_-}}\!$}} 

\bibnote{2011ApJ...736...71W}{(Paper~I)} 
\def\pI{\href{http://adsabs.harvard.edu/abs/2011ApJ...736...71W}{Paper~I}}
\bibnote{2013ApJ...774...32V}{(Paper~II)} 
\def\pII{\href{http://adsabs.harvard.edu/abs/2013ApJ...774...32V}{Paper~II}}
\bibnote{2015ApJ...812...11V}{(Paper~III)} 
\def\pIII{\href{http://adsabs.harvard.edu/abs/2015ApJ...812...11V}{Paper~III}}
\bibnote{2015ApJ...808..133R}{(Paper~IV)} 
\def\pIV{\href{http://adsabs.harvard.edu/abs/2015ApJ...808..133R}{Paper~IV}}

\begin{document}

\title{Reconnection brightenings in the quiet solar photosphere}

\author{Luc H. M. Rouppe van der Voort\inst{1} 
\and
Robert J. Rutten\inst{1,2} 
\and
Gregal J. M. Vissers\inst{1}
}

\institute{Institute of Theoretical Astrophysics,
  University of Oslo, %
  P.O. Box 1029 Blindern, N-0315 Oslo, Norway
\and
Lingezicht Astrophysics, 't Oosteneind 9, 4158\,CA Deil, The Netherlands}

\abstract{%
We describe a new quiet-Sun phenomenon which we call ``Quiet-Sun
Ellerman-like Brightenings'' (QSEB). 
\acp{QSEB}s are similar to Ellerman bombs (EB) in some respects but
differ significantly in others.
\acp{EB}s are transient brightenings of the wings of the Balmer
\Halpha\ line that mark strong-field photospheric reconnection in
complex active regions.
\acp{QSEB}s are similar but smaller and less intense Balmer-wing
brightenings that occur in quiet areas away from active regions. 
In the \Halpha\ wing we measure typical lengths of less than 
0.5~arcsec,
widths of 0.21~arcsec, and lifetimes of less than a minute.
We discovered them using high-quality \Halpha\ imaging spectrometry
from the Swedish 1-m Solar Telescope (SST) and show that in
lesser-quality data they cannot be distinguished from more ubiquitous
facular brightenings, nor in the ultraviolet diagnostics currently
available from space platforms. 
We add evidence from concurrent \acp{SST} spectropolarimetry that
\acp{QSEB}s also mark photospheric reconnection events, but in quiet
regions on the solar surface. 
}%

\maketitle

\section{Introduction}
\label{sec:introduction}

Reconnection brightenings in the solar photosphere are well-known in
their manifestation as Ellerman bombs (henceforth EB).
\acp{EB}s are small but intense brightenings of the extended wings of the
hydrogen Balmer-$\alpha$ line at 6563~\AA\ (henceforth \Halpha) that
occur intermittently in complex bipolar active regions and mark
magnetic reconnection in the low solar atmosphere. 

Here we report the discovery of similar but weaker \Halpha-wing
brightenings in quiet solar areas away from active regions.
We call them ``quiet-Sun Ellerman-like brightenings'' (henceforth
\acp{QSEB}) and show evidence that also these mark magnetic reconnection in
the low solar atmosphere.

\citetads{1917ApJ....46..298E} 
described his ``hydrogen bombs'' so well that in their review
\citetads{1974soch.book.....B} 
concluded that the subsequent \acp{EB} literature at the time did not add
much to his description.
More recently this situation changed, first with high-quality
observations from the Flare Genesis balloon telescope
(\citeads{1999ASPC..183..279B}; 
\citeads{2002ApJ...575..506G}; 
\citeads{2004ApJ...601..530S}; 
\citeads{2004ApJ...614.1099P}; 
\citeads{2006AdSpR..38..902P}) 
and then with yet better observations from the Swedish 1-m Solar
Telescope (\citeads[SST,][]{2003SPIE.4853..341S}; 
\citeads{2011ApJ...736...71W}, 
henceforth \pI; %
\citeads{2013ApJ...774...32V}, 
henceforth \pII; %
\citeads{2015ApJ...812...11V}, 
henceforth \pIII; %
\citeads{2015ApJ...808..133R}, 
henceforth \pIV; %
\citeads{2015ApJ...798...19N}; 
Reid et al.\ \citeyearads{2015ApJ...805...64R}, 
\citeyearads{2016arXiv160307100R}). 

These studies have firmly established that \acp{EB}s result from
strong-field reconnection taking place in the photosphere.  
They contain so much heating that they become visible
even in ultraviolet lines normally arising from the transition region
(\pIII; 
 \citeads{2016arXiv160405423T}). 

In observing \acp{EB}s near solar disk center it is nontrivial to
distinguish them from ubiquitous strong-field magnetic concentrations
(henceforth \acp{MC}) that are abundant in active regions and constitute
plage and network elsewhere. 
These also brighten the \Halpha\ wings
(\citeads{2006A&A...449.1209L}). 
Many papers in the \acp{EB} literature have confused such \acp{MC} brightenings
with \acp{EB} brightenings (``pseudo EBs'',
\citeads{2013JPhCS.440a2007R}). 
Therefore, in Papers I--IV our strategy has been to observe \acp{EB}s well
away from disk center.
In slanted-view \Halpha-wing images \acp{EB}s appear as ``flames'' with
characteristic morphology: upright, tall (1\,--\,2\,Mm), rapidly
flickering on fast time scales (seconds), and often with their feet
traveling at high speed (1\,\kms) along network lanes filled with \acp{MC}s
and then occurring repetitively in quick succession (\pI).   
These signature properties became our way to distinguish \acp{EB}s from \acp{MC}s
in high-resolution imaging spectrometry with the \acp{SST} (\pII, \pIII, \pIV).

We employed the same strategy of limbward flame detection when we
decided to search for \acp{EB}-like phenomena in quiet areas outside active
regions. 
This search was inspired by numerical simulations presented by
S.~Danilovic at a recent ISSI meeting in Bern. 
In her simulations reconnection in the quiet low photosphere produced
small flames in synthesized \Halpha-wing images that appear rather
like observed \acp{EB}s, although smaller. 
We found that such small \acp{EB}-like flames indeed occur also outside
active regions on the actual solar surface, but that their detection is only
possible at the very best seeing at the \acp{SST} and even then requires the
utmost of the advanced adaptive optics and image restoration developed
at this telescope.

\begin{table*}[bth]
\caption{Overview of the CRISP datasets analyzed in this study.}
\centering
\begin{tabular}{ccccccccccc}%
        \hline \hline
        Set & Date  
& Pointing & $\mu$\tablefootmark{1} & Time & Duration & Range\tablefootmark{2}  & Sampling & Cadence & Other  \\
        {} & {} 
& (solar $X,Y$) & {} & (UT) & [h:m:s] & [\AA] & [m\AA] & [s] & instruments \\
        \hline
        1 & 2013 Jul 4 
& $-96$\arcsec, 931\arcsec & 0 -- 0.33 & 09:20:32 & 00:21:45 & $\pm 1.400$ & 200 & 3.3 & {} \\
        2 & 2013 Jul 4 
& 914\arcsec, $-127$\arcsec & 0 -- 0.33 & 10:13:55 & 00:53:00 &  $\pm 2.064$ & 84\tablefootmark{3} & 8.6 & AIA \\
        3 & 2014 Jun 17 
& 24\arcsec,939\arcsec & 0 -- 0.30 & 10:20:03 & 00:55:49 & $\pm 1.200$ & 100 & 5.5 & AIA, IRIS \\
        4 & 2014 Jun 19 
& $-11$\arcsec, 797\arcsec & 0.56 & 07:27:10 & 00:53:52 & $\pm 1.200$ & 100 & 5.5 &  \\
        5 & 2014 Jun 19 
& $-11$\arcsec, 891\arcsec & 0 -- 0.38 & 08:26:24 & 01:19:30 & $\pm 1.200$ & 100 & 5.5 & AIA \\
        6 & 2014 Sep 22 
& 378\arcsec,$-412$\arcsec & 0.81 & 08:37:41 & 02:16:35 & $\pm 1.400$\tablefootmark{4,5} & 200 & 11.7 & \\
        7 & 2015 Sep 15 
& 652\arcsec, $-47$\arcsec & 0.73 & 09:24:58 & 01:21:32 & $\pm 1.400$\tablefootmark{4,6} & 200 & 24.1 & AIA, IRIS \\%
        8 & 2015 Sep 17 
& 728\arcsec, 44\arcsec & 0.64 & 09:02:45 & 00:24:31 & $\pm 1.400$\tablefootmark{4,6} & 200 & 24.1 & AIA, IRIS  \\
        9 & 2015 Oct 11 
& $-591$, 52\arcsec & 0.79 & 08:44:16 & 01:57:16 & $\pm 1.400$\tablefootmark{4,6} & 200 & 24.1 & AIA, IRIS \\
        \hline
\end{tabular}
\tablefoot{%
\tablefoottext{1}{$\mu = \cos \theta$ with $\theta$ the observing angle.}
\tablefoottext{2}{Extent of spectral coverage of \Halpha.}
\tablefoottext{3}{Regular 84~m\AA\ sampling over $\pm 1.376$~\AA, 
and further out at $\pm 1.548, \pm 1.806$ and $\pm 2.064$~\AA.}
\tablefoottext{4}{In addition to \Halpha\ spectral scans of 
\CaIR\ were included, with regular 100~m\AA\ sampling 
over $\pm 1.2$~\AA.}
\tablefoottext{5}{A single-wavelength polarimetric sampling of 
\ion{Fe}{i}~6302~\AA\ at $-48$~m\AA\ was included to yield Stokes I, Q, U, and V maps.}
\tablefoottext{6}{A 12-wavelength polarimetric sampling of \FeIline\ was 
included with regular 35~m\AA\ sampling over $\pm 0.150$~\AA, 
further out at $\pm 0.240$ and $+0.315$~\AA\ (continuum).}
}
\label{table:datasets}
\end{table*}%

This decisive need for highest image quality results from the
likelihood of confusion between \acp{QSEB}s and \acp{MC}s, similar to the
confusion between \acp{EB}s and \acp{MC}s nearer disk center. 
While \acp{MC}s constituting network appear as roundish \Halpha-wing
brightness features in top-down viewing, they become elongated towards
the limb in becoming faculae.
This is because in top-down viewing the \Halpha-wing brightening is
from subsurface hot-wall emission
(\citeads{1976SoPh...50..269S}; 
\citeads{2006A&A...449.1209L}), 
whereas in limbward observation relatively empty fluxtubes are mapped
by permitting deeper slanted viewing into hot granule interiors behind
them (cartoon in Fig.~7 of
\citeads{1999ASPC..184..181R}; 
simulations in \citeads{2004ApJ...607L..59K} 
and \citeads{2004ApJ...610L.137C}). 
The resulting facular ``striations'' have elongated upright stalk
appearance and so may masquerade as small reconnection flames
(``pseudo QSEBs'').
However, at the image quality of the \acp{SST} during the best La Palma
seeing definite distinction can be made, as we demonstrate below.  

The next section presents the observations and their reduction.
We then present the results and end the paper with discussion and
conclusion.

\begin{figure*}
\centerline{%
\includegraphics[width=\columnwidth]{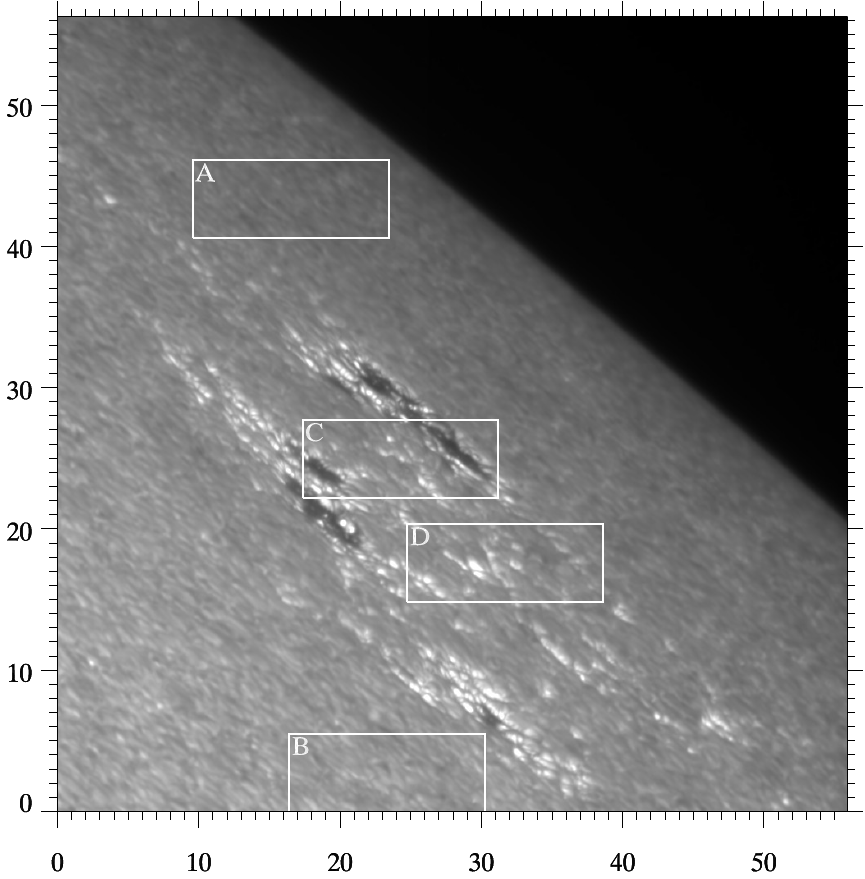}
\includegraphics[width=\columnwidth]{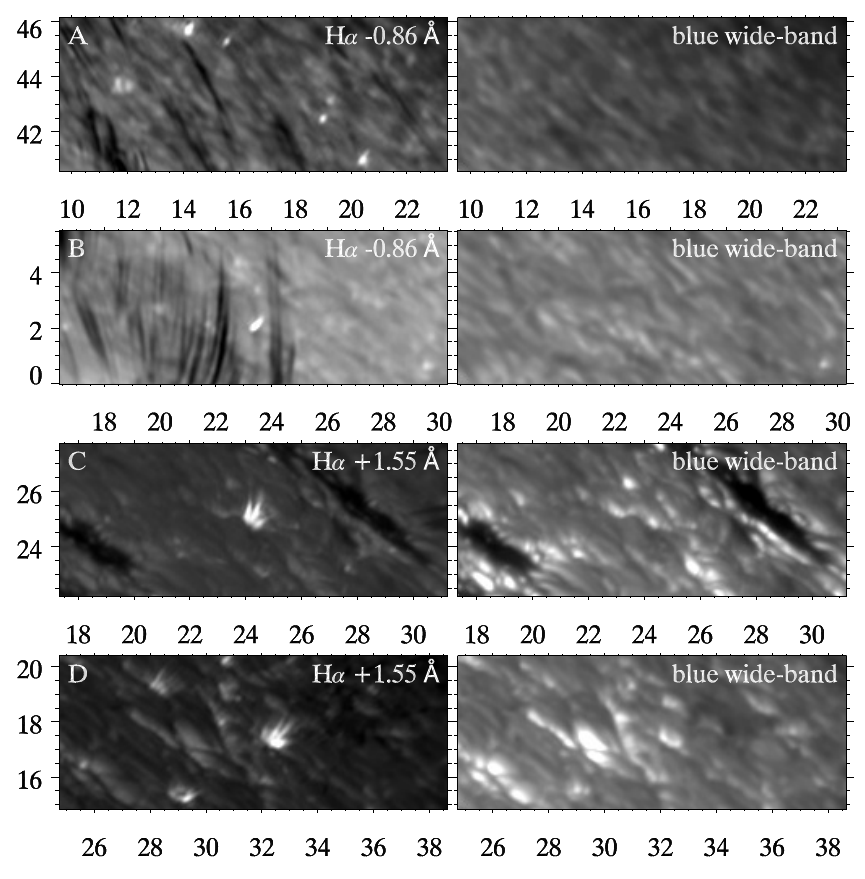}}
\caption{\label{fig:overview}%
Examples of \acp{EB}s and \acp{QSEB}s in dataset~2 which targeted active region
AR~11778 near the limb.
The \acp{SST} field is rotated over $-59$~degrees from solar $(X,Y)$.
The full-field overview at left was taken with the blue
($\lambda=3953.7$~\AA) wide-band (FWHM 10~\AA) camera. 
The axis divisions are arcsec from the lower-left corner.
The four regions of interest A--D are shown enlarged as \Halpha-wing
and blue wide-band cutout pairs at right, with corresponding arcsec
coordinates. 
Regions A and B are in quiet areas, regions C and D sample the active
region. 
Ellerman features are visible as bright flames in the \Halpha\ wings,
respectively as smaller and weaker \acp{QSEB}s in A and B and as larger and
brighter \acp{EB}s in C and D.
Their relative brightness enhancements can be judged from the darkness
of the surrounding granulation because each cutout panel is
independently greyscaled. 
They are invisible in the blue wide-band images at right, in contrast
to facular striations which are visible in both diagnostics, although
appearing darker in \Halpha\ from the greyscaling needed to
accommodate the bright Ellerman flames.
}
\end{figure*}

\begin{figure*}
\centerline{\includegraphics[width=\textwidth]{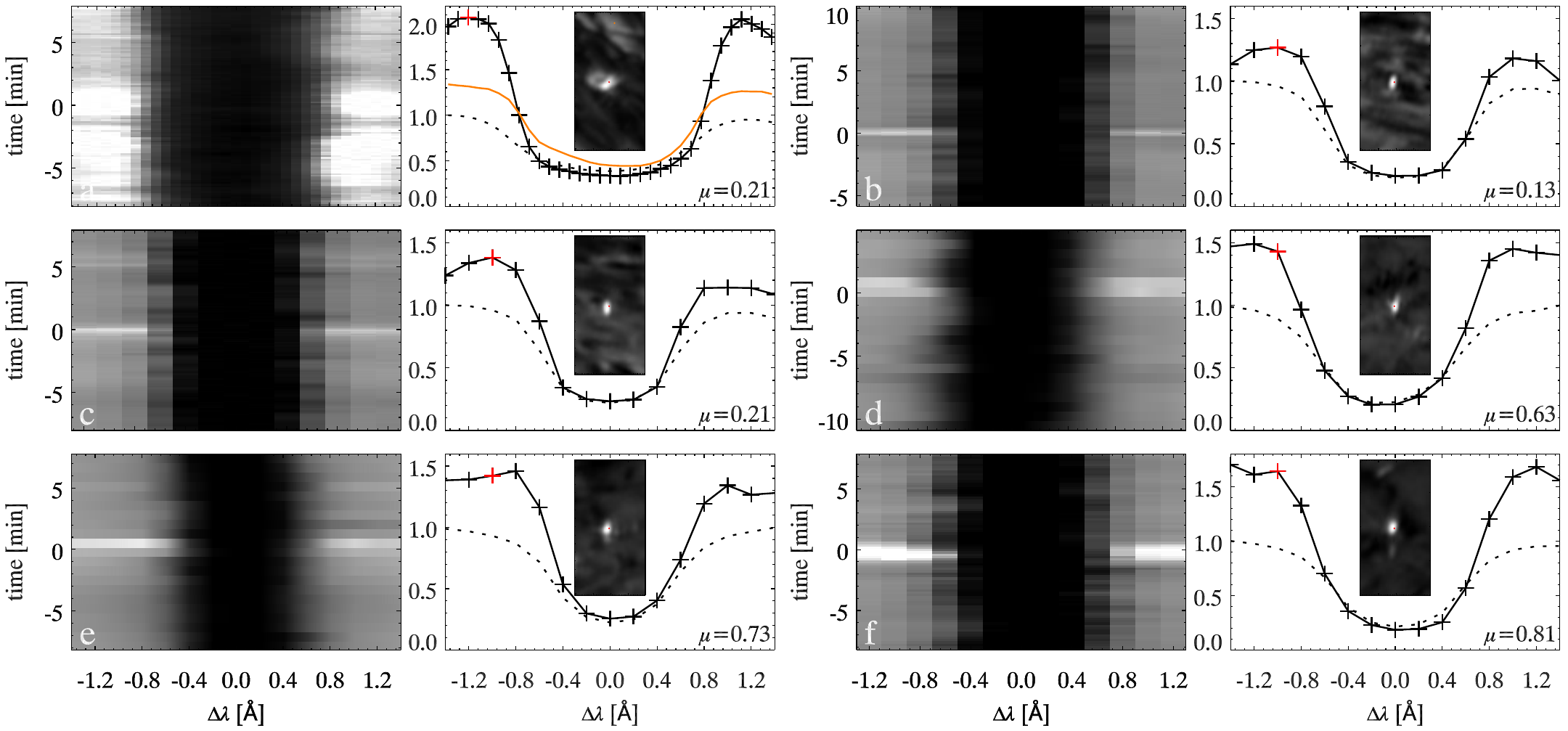}}
\caption{\label{fig:lamt}%
Sample $\lambda-t$ diagrams and spectral profiles of Ellerman features
in \Halpha\ at the observing angles specified at lower-right in the
profile plots.
Panel pair $(a)$ is from dataset 2, $(b)$ and $(c)$ from 1, $(d)$ from
8, $(e)$ from 7, $(f)$ from 6.
Pair $(a)$ shows an active-region \acp{EB} for comparison, the one
at the bottom at $(x,y)\approx(29\arcsec,15\arcsec)$ 
in \Halpha-panel D of Fig.~\ref{fig:overview}.
The other pairs show \acp{QSEB}s.
Each $\lambda-t$ image is greyscaled between 0.3 and 1.5 times the
\Halpha\ far wing intensity.
The Ellerman feature occurs at time 0. 
The profiles are normalized to the outer-wing intensity of the
time-averaged profile (dashed) of an area close to the Ellerman
feature.
Note the scale difference for profile graph $(a)$.
The orange profile added to $(a)$ is from the facular brightening
visible at the top of the small inset image.
The red pluses in the blue wing of each feature profile (solid)
specify the spectral sampling of the small image insets 
(size 2.3 by 4.6~arcsec)
in which central red dots mark the location of the
spectral sampling. 
In pair $(a)$ the location of the facular sampling is also marked in
the inset (easier seen per zoom-in with a pdf viewer). 
The limb direction is upward.
}
\end{figure*}

\begin{figure*}[bht]
\centerline{\includegraphics[bb=0 0 482 75, clip, width=\textwidth]{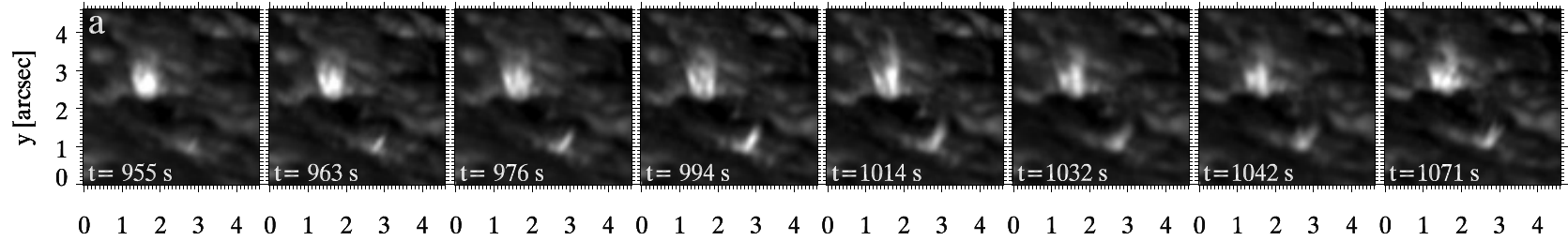}}
\centerline{\includegraphics[bb=0 13 482 75, clip, width=\textwidth]{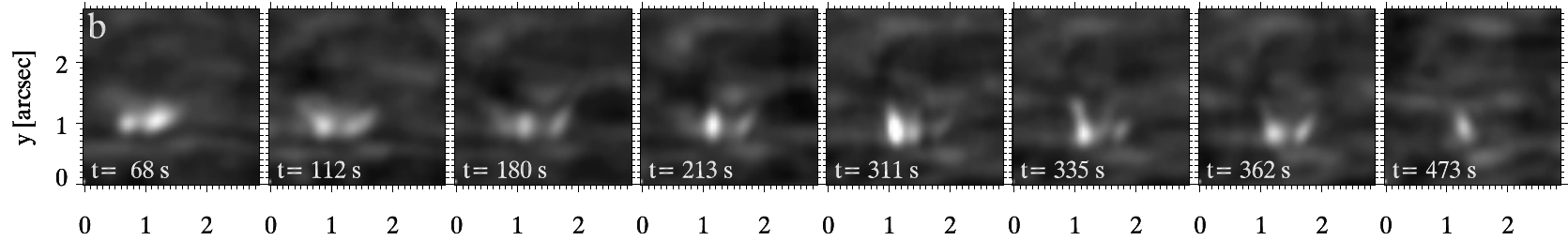}}
\centerline{\includegraphics[bb=0 13 482 75, clip, width=\textwidth]{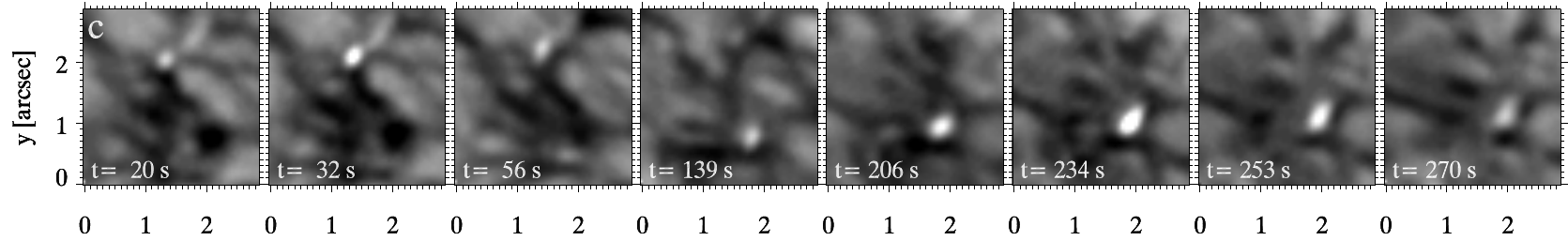}}
\centerline{\includegraphics[bb=0 0 482 75, clip, width=\textwidth]{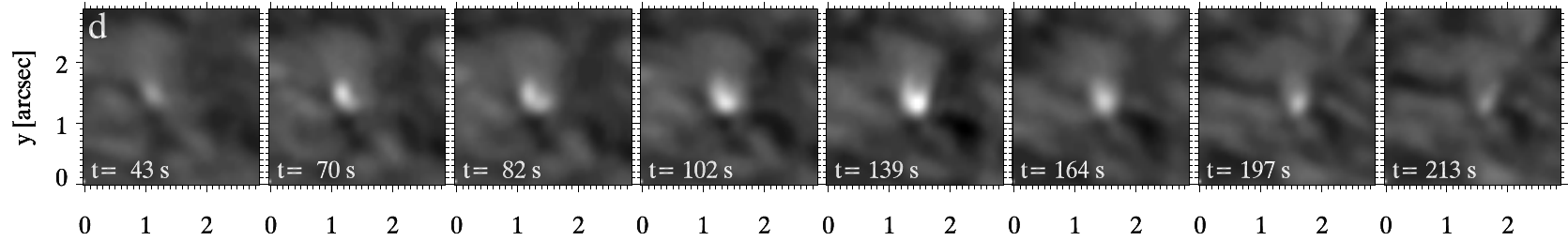}}
\caption{\label{fig:wbfast}%
Temporal evolution of Ellerman features at 1-s cadence in \Halpha\
wide-band (FWHM=4.9~\AA) image sequences. 
Each row of image cutouts shows selected samples with identical
greyscaling per panel.
The limb direction is upward. 
Row $(a)$ contains an active-region EB from dataset 2 (close to the limb, cutout C in
Fig.~\ref{fig:overview}).
The other rows show \acp{QSEB}s in quiet areas in yet smaller cutouts from
datasets 5 ($b$, close to the limb) and 4 ($c$ and $d$, $\mu =0.56$).
The time labels specify duration from the start of each observation. 
Corresponding animations of this figure are available at 
\url{http://folk.uio.no/rouppe/qseb/}.
}
\end{figure*}
%

\begin{figure*}[bht]
\centerline{\includegraphics[width=\columnwidth]{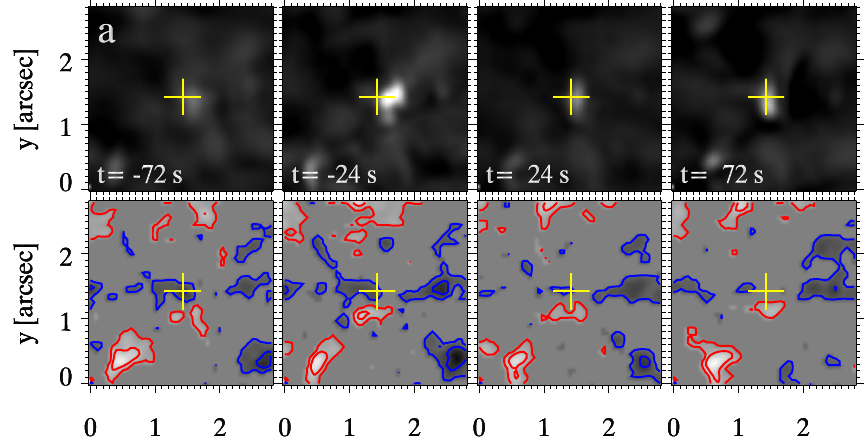}
              \includegraphics[width=\columnwidth]{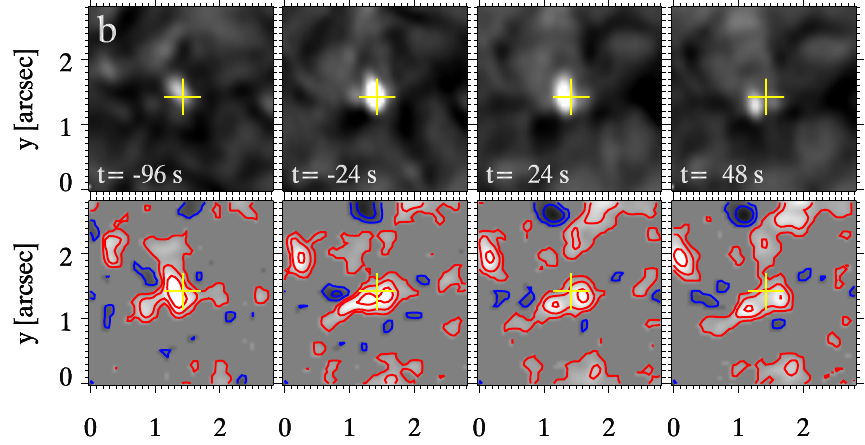}}
\centerline{\includegraphics[width=\columnwidth]{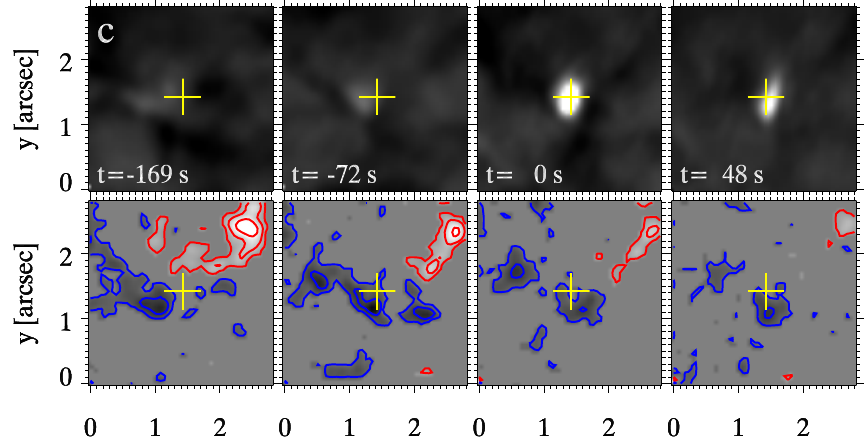}
              \includegraphics[width=\columnwidth]{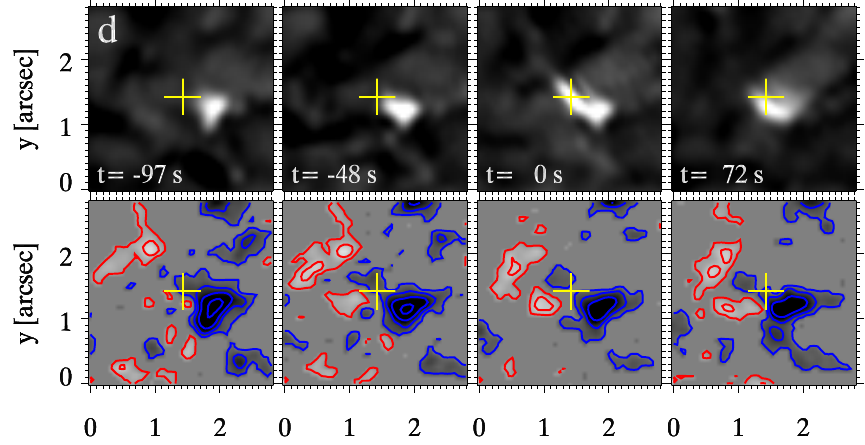}}
\caption{\label{fig:stokes}%
Four examples of magnetic field evolution around \acp{QSEB}s. 
The four upper panels per sequence show \Halpha\ wing image cutouts at
$\Delta \lambda=-1$~\AA, the four lower panels corresponding Stokes
$V/I_{\rm cont}$ maps obtained from CRISP spectropolarimetry in \FeIline.
These are averages over 6 wing positions at
$\Delta \lambda=\pm 0.150$, $\pm 0.105$, and $\pm 0.070$~\AA\ from
line center. 
Signals below 3$\sigma_{\rm noise}$ are put to 0 in the greyscaling,
positive polarity is white with red contours, negative polarity is
black with blue contours. 
The contour levels are at 0.2, 0.5, 1, and 2\% polarization. 
The yellow cross at the center of each panel serves to guide
inspection. 
$(a)$ and $(c)$ are from dataset 7 ($\mu=0.73$), $(b)$ and $(d)$ from dataset 8 ($\mu=0.64)$. 
Animations of this figure are available at 
\url{http://folk.uio.no/rouppe/qseb/}.
}
\end{figure*}
%

\begin{figure*}
\sidecaption
\includegraphics[width=120mm]{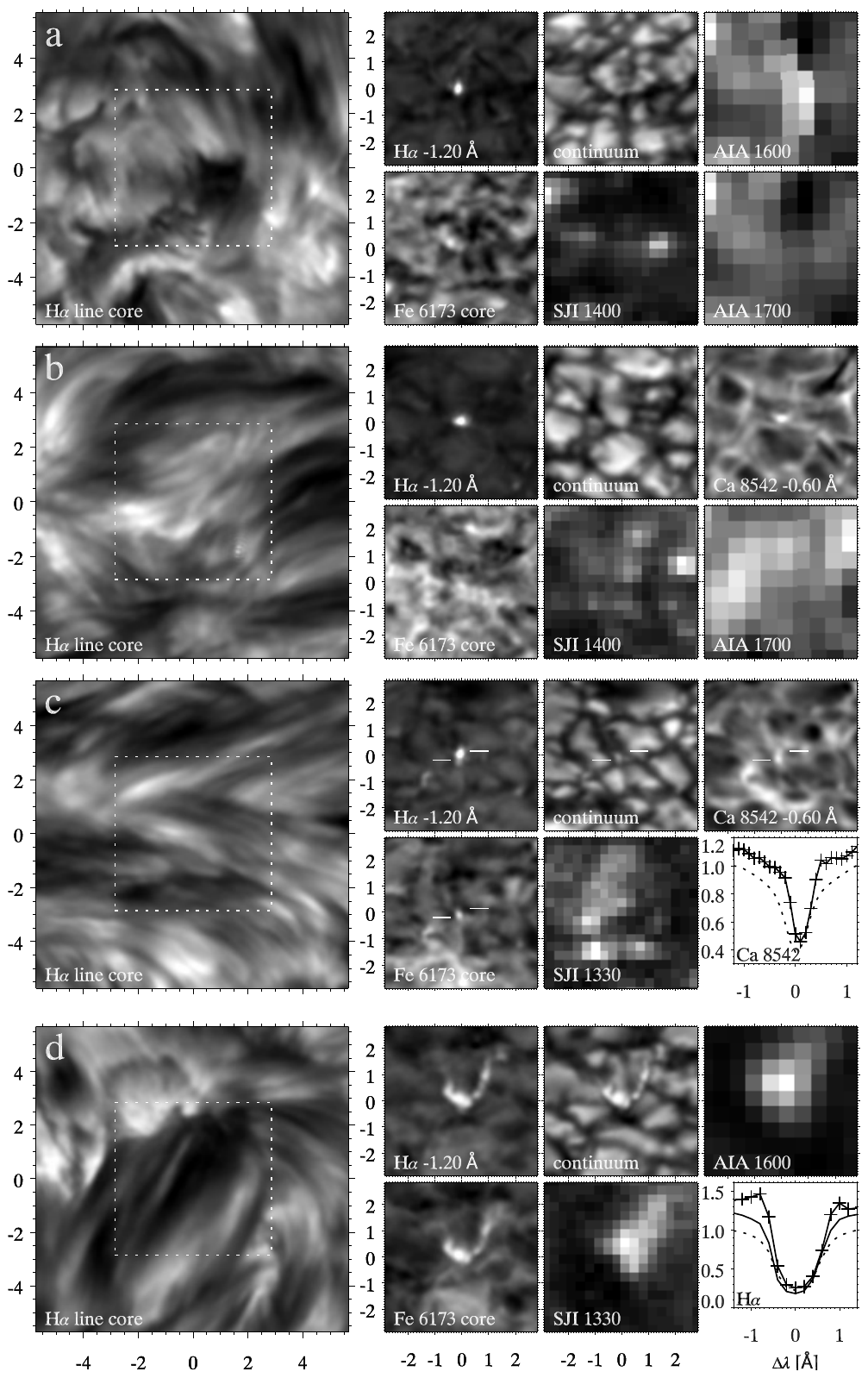}
\caption[]{\label{fig:diag}%
Three \acp{QSEB}s and one pseudo-QSEB in \Halpha\ line center over a fairly
wide field of view (first column) and in smaller image cutouts
clockwise sampling the blue wing of \Halpha, the wide-band continuum
around \Halpha, AIA 1600\,\AA\ or \CaIR\ wing, AIA 1700\,\AA, IRIS slit-jaw 1400\,\AA\
or 1330\,\AA, and the core of \FeIline.
The dotted frame in the first column specifies the smaller cutouts at
right.
The axis units are arcsec from the feature at $(0,0)$.
The first row is from dataset 8 ($\mu=0.64$), 
the others are from dataset 7 ($\mu=0.73$).
The \acp{QSEB} in row $(c)$ is the same as in Fig.~\ref{fig:lamt}\,$(e)$;
for this one the AIA 1700 image is replaced by a \CaIR\ profile plot.
White horizontal markers are added to guide the eye: the
\acp{QSEB} visible in the \Halpha\ wing is slightly above the small
bright point also visible in the other \acp{SST} diagnostics (marked with
the lower left marker).
Row $(d)$ shows a pseudo-QSEB, i.e., facular striations.
The AIA 1700\,\AA\ image is replaced by a plot of the \Halpha\ profile
of the facula at image center (solid profile without crosses).
The other profiles (with crosses and dashed) are for the \acp{QSEB}
and its reference spectrum in row $(c)$, copied from
Fig.~\ref{fig:lamt}\,$(e)$.
\vspace{2ex}\mbox{} 
}
\end{figure*}

\section{Observations and data reduction}
\label{sec:observations}

Over the past decade the Oslo group has collected over 70 high-quality
datasets with the \acp{SST}. 
From these we have selected the very best in our search for limbward
Ellerman-like phenomena in quiet areas.
The ones from which we present results here are from the 2013, 2014,
and 2015 observing seasons and are detailed in
Table~\ref{table:datasets}. 
They targeted quiet areas, except for the second dataset in which the
field of view contained an active region (Fig.~\ref{fig:overview}).
Datasets 1--3 and 5 included the limb in the field of view (see $\mu$
specification in the fourth column).

These data were obtained with the CRISP Fabry-P\'erot interferometer
at the \acp{SST} \citepads{2008ApJ...689L..69S}. 
It provides imaging spectroscopy in \Halpha\ with transmission profile
FWHM 66~m\AA, image scale 0\farcs058 per pixel, and a field of view of
about 60\arcsec~$\times$~60\arcsec.
Some datasets also contain CRISP spectral imaging in \CaIR\ with FWHM
107~m\AA\ and imaging spectropolarimetry in \FeIline\ with FWHM
51~m\AA.

\acp{QSEB}s are small and can only be identified in \acp{SST} data that reach the
telescope diffraction limit $\lambda/D$ = 0\farcs14 at 6563\,\AA.
Imaging of this quality requires not only a superb telescope
and extremely stable atmospheric conditions, but also 
high-order adaptive optics,
at the \acp{SST} employing an 85-electrode deformable mirror, and advanced
subsequent image restoration, in our case multi-object multi-frame
blind deconvolution
\citepads[MOMFBD,][]{2005SoPh..228..191V}. 
In its application here we used eight exposures per CRISP spectral
line position (or spectropolarimetric state for \FeIline).
We followed the extensive CRISPRED data-reduction pipeline
(\citeads{2015A&A...573A..40D}) 
including 
time-dependent derotation, coalignment and destretching.

For the \Halpha-only CRISP datasets 1--6 we also processed the
\Halpha\ prefilter data (FWHM 4.9~\AA) in parallel in order to obtain
simultaneous sequences sampling the low photosphere at a cadence of
only 1\,s. 
These permit to study fast temporal evolution within \acp{QSEB}s similarly
to the \acp{EB} analysis in \pI\ (Fig.~\ref{fig:wbfast}).

For datasets 1 and 2 we include imaging with the \acp{SST} blue beam that
consist of MOMFBD-restored filtergrams in \ion{Ca}{ii}~H~3969~\AA\
(FWHM 1.1~\AA), \ion{Ca}{ii}~K~3934~\AA\ (FWHM 1.5~\AA), and with a
wider ``blue'' passband (FWHM 10~\AA) at $\lambda=3954$~\AA\ between
the \ion{Ca}{ii}~H and K lines which shows the lower photosphere.
The image scale for these is 0\farcs034; the field of view slightly
exceeds the CRISP imaging.
They were recorded at 10.8 frames per second with 9.5~ms exposure and
MOMFBD-restored to yield sequences at the cadence of the cotemporal
CRISP data (3.3 and 8.6~s, see Table~\ref{table:datasets}).

To check for \acp{QSEB} signatures in hotter diagnostics sampling the upper
chromosphere, transition region and corona for some datasets we also
coaligned cotemporal observations from the Interface Region Imaging
Spectrograph 
\citepads[IRIS,][]{2014SoPh..289.2733D} 
and from the Atmospheric Imaging Assembly \citepads[AIA,][]
{2012SoPh..275...17L} 
onboard the Solar Dynamics Observatory (SDO,
\citeads{2012SoPh..275....3P}), 
as specified in Table~\ref{table:datasets}.

From IRIS, we analyzed slit-jaw images (SJI) in the 1400 and 1330~\AA\
channels (dominated by \ion{Si}{iv} and \ion{C}{ii} lines,
respectively, both filters have FWHM 55~\AA), and the 2796~\AA\
channel containing \ion{Mg}{ii}~k in its FWHM 4~\AA\ passband.
For datasets 7--9 the slit-jaw data had 12~s cadence, 2~s exposure
time, and spatial binning over 0\farcs33 pixels.
The alignment to the \acp{SST} data was done through cross-correlation of
the SJI 2796~\AA\ images to the CRISP images in the \CaIR\ wing.
In quiet solar areas they show very similar scenes. 
The accuracy of this alignment is about the IRIS pixel size.
Dataset 3 has only SJI 1400 and 2796~\AA\ at 19~s cadence, 8~s exposure
time and no spatial binning (0\farcs17 pixels). 
For this limb dataset, alignment to the \acp{SST} data was done through cross-correlation of the on-disk part of
the SJI 2796~\AA\ and \ion{Ca}{ii}~H images.

The alignment of the AIA data, which have an image scale of 0\farcs6
per pixel, was done for dataset 3 
through cross-correlation of 
AIA 1600~\AA\ images to \ion{Ca}{ii}~H
images, for dataset 9 by aligning continuum images from the
Helioseismic and Magnetic Imager
\citepads[HMI,][]{2012SoPh..275..229S} 
to CRISP \Halpha\ wing images, and for the other datasets by aligning
HMI continuum images to CRISP \Halpha\ prefilter images.  

For our data searches and our successful identifications of \acp{QSEB}s we
made extensive use of the CRISPEX imaging spectroscopy viewer of
\citetads{2012ApJ...750...22V}, 
a widget-based IDL tool for efficient inspection and analysis of
multi-dimensional spectral-imaging data. 
It is available in the SolarSoft library as part of the IRIS package.

\section{Results}   \label{sec:results}

\paragraph{Morphology.}
The defining appearance and morphology of \acp{QSEB}s as small bright
\Halpha-wing flames is illustrated in Fig.~\ref{fig:overview},
including comparison with larger and brighter \acp{EB} flames. 
The field of view of this dataset (nr.~2) is shown at left and
includes a young emerging active region close to the limb as well as
extended quiet areas.
The righthand panels enlarge two cutouts in quiet areas (A and B) and
two comparison ones in the active region (C and D). 

The \Halpha\ red-wing images for C and D contain prominent \acp{EB}s. 
They appear as upright flames quite similar to the \acp{EB}s in \pI, with
the tallest in C nearly reaching 1\farcs5 (1.1~Mm) in length and
adhering to the remark by \citetads{1917ApJ....46..298E} 
that his bombs typically occur between spots in complex developing
active regions. 

The blue wide-band images at right do not show the \acp{EB}s in C and D but
instead bright faculae bordering granular "hills" seen from aside
(\citeads{2004ApJ...607L..59K}; 
\citeads{2004ApJ...610L.137C}). 
All faculae are also visible in the \Halpha-wing images, in which
their contrast over the granulation is actually larger (see last
panels of Fig.~\ref{fig:diag}), but here they appear less
conspicuous because the greyscale is stretched to include the far
brighter \acp{EB}s.
The marked differences in contrast and morphology confirm that \acp{EB}s are
distinct from facular striations, even though also these appear
elongated and generally upright in the limbward projection.

The quiet regions in A and B show \acp{QSEB}s in the inner blue \Halpha\
wing as small elongated brightenings. 
They are comparable to the \acp{EB}s in C and D but less tall and less
bright. 
Their invisibility in the blue companion images at right demonstrates
that these quiet-Sun features are not faculae, just as for the
active-region \acp{EB}s in C and D.
The tallest example is the \acp{QSEB} in panel B.
It is visible through a forest of dark spicular features in the
foreground which are rapid blue-shifted excursions (RBE,
\citeads{2009ApJ...705..272R}). 
This \acp{QSEB} has length 0\farcs6 (0.4~Mm) and width below 0\farcs25
(0.18~Mm). 
Panel A contains multiple smaller \acp{QSEB}s, with the smallest hardly
elongated and reaching only 0\farcs29 (0.21~Mm) extent, marginally
taller than its width of 0\farcs21 (0.15~Mm). 
This sample is closer to the limb than B, causing darker granulation
background at similar \acp{QSEB} brightness. 
 
From similar measurements for a sample of 24 \acp{QSEB}s in dataset~1
and 21 \acp{QSEB}s in dataset~2 we conclude that \acp{QSEB}s typically
have lengths below 0\farcs5 (0.36~Mm) and widths of only 0\farcs23
(0.17~Mm). 
We selected these limb datasets for these measurements on the
assumption that \acp{QSEB}s generally stand upright, so that near-limb
viewing effectively gives a side view minimizing projection
uncertainty in the length measurement. 
The typical lifetimes for these \acp{QSEB}s was measured to be less
than a minute. 

\paragraph{Spectral detail.}
Figure~\ref{fig:lamt} continues our presentation of \acp{QSEB}s by showing
spectral evolution and detailed line profiles of \Halpha\ for multiple
examples.

The first pair of panels is not a \acp{QSEB} but an active-region \acp{EB} serving
as comparative reference. 
Its \Halpha\ profile has enhanced wings that reach maximum at
$-1.2$~\AA\ from line center and double the far-wing intensity of the
local reference spectrum. 
Near line center, the \acp{EB} spectral profile is as dark as the reference
due to obscuration by dark chromospheric fibrils overlying the
photospheric reconnection site below the canopy (\pI). 
The spectral evolution slice at left shows that this \acp{EB} was preceded
by multiple such flarings during the preceding minutes, a common \acp{EB}
characteristic that was already noted by
\citetads{1917ApJ....46..298E}. 

The orange profile in the graph is not for an Ellerman feature but
from the facular \acp{MC} at the top of the small image inset. 
It demonstrates that faculae brighten the \Halpha\ wings less than \acp{EB}s
do.

Panel pairs $(b)-(f)$ of Fig.~\ref{fig:lamt} show example \acp{QSEB}s
ordered over increasing viewing angle $\mu$. 
They show similarly enhanced \Halpha\ wings, but with lower contrast
than the \acp{EB} in pair $(a)$.
However, some yet reach maximum \Halpha-wing intensity equal to or
exceeding the value of 1.5 times the reference far-wing intensity
which was defined as intensity threshold for automated \acp{EB} detection in
\pII. 
But even when \acp{QSEB}s reach this \acp{EB} threshold we would not designate them
as \acp{EB}s since they are not in active regions as described by
\citetads{1917ApJ....46..298E}. 
In addition, they have discordant spectral visibilities described
below.

Examples $(b)$ and $(c)$ stay below 1.5 and have wing intensities
comparable to the orange facular profile in $(a)$.
However, their spectral evolutions in the $\lambda-t$ diagrams show
brief lifetimes, below a minute, while faculae evolve at granular time
scales and last multiple minutes. 
We also note that faculae as bright as the orange profile in $(a)$ are
mostly found in active-region plage and strong network, rarely in
quiet-Sun areas.
Furthermore, the \acp{QSEB} spectral profiles are different from faculae:
they taper off towards larger offset beyond their maximum at about
$-1$~\AA, while the facular profile keeps rising towards the blue
limit of the CRISP scan (this can only be inspected for the blue
\Halpha\ wing; the red wing decreases for both \acp{QSEB}s and faculae from
line blends around $+1.3$~\AA).
Indeed, faculae also brighten the continuum while \acp{QSEB}s do not
(Fig.~\ref{fig:diag}). 

Examples $(e)$ and $(f)$ are from fields nearer disk center and 
indeed appear more roundish than flame-like.  Their wing brightness
reaching 1.5 is the main discriminator from faculae.

In all profile graphs the line core of the feature duplicates the core
of the reference profile. 
The reason is that in all areas the \Halpha\ core originates in
obscuring overlying fibrils that together constitute the chromospheric
canopy. 
Examples are shown in the first column of Fig.~\ref{fig:diag} 
(in particular, the \Halpha\ core image of the \acp{QSEB} from $(e)$ is shown in Fig.~\ref{fig:diag}$c$).

Note that none of these \acp{QSEB}s showed repetitive flaring over the
course of minutes as was the case for the \acp{EB} in pair $(a)$ and happens
often for \acp{EB}s.

\paragraph{Temporal behavior.}
Figure~\ref{fig:wbfast} illustrates temporal behavior of an \acp{EB} (top
row) and multiple \acp{QSEB}s (other rows).   
The images are taken with a wide-band \Halpha\ filter and are part of
several time series that have a temporal cadence of 1~s. 
Figure~\ref{fig:wbfast} shows eight selected images from each time
series, the corresponding animations are available in the online
material. 
The high cadence reveals fast dynamical evolution in these events,
similar to the  \acp{EB}s in similar \Halpha\ wide-band data in \pI.

The 21-min \acp{EB} movie corresponding to the top row of
Fig.~\ref{fig:wbfast} shows complex dynamical evolution of \acp{EB}s at
several sites in this area.
The site in the top left quadrant has \acp{EB}s present throughout the full
movie duration.
There seem to be both upflows and downflows associated with these \acp{EB}s,
with apparent downflows mostly in the bottom part and apparent upflows
mostly in the top.

The three \acp{QSEB} movies show events that are less tall but display
similar morphology and similar fast changes with time, faster than the
changes in the background granulation.
The darker background of the \acp{EB} images illustrates that its brightness
is larger than the \acp{QSEB} brightnesses. 
During the $\sim$9 min duration of the \acp{QSEB} sequence $(b)$, one can
distinguish at least six different \acp{QSEB} events. 
Some show comparable up and downflows as in the \acp{EB} sequence. 
The lower two sequences are further from the limb; the \acp{QSEB}s show less
clear upright-flame morphology than in the top \acp{QSEB} sequence. 
Sequence $(c)$ shows two separate, short-lived \acp{QSEB} events while
the bottom sequence shows a \acp{QSEB} with signs of the fast repetitive
flaring that is so clearly present in some of the \acp{EB} events. 
  
\paragraph{Magnetic surroundings.}
Figure~\ref{fig:stokes} compares \acp{QSEB} evolution for four cases with
the magnetic environment obtained from \FeIline\ polarimetry.  
In each case the topography is bipolar, with the ``pepper-and-salt''
patterning that is characteristic of quiet areas.  
In all, the \acp{QSEB} originates close to one fairly strong patch with
comparable opposite polarity patches not far away.
Case $(a)$ suggests cancellation of negative polarity against opposite
polarity at the \acp{QSEB} site; both polarities diminish in presence. 
Case $(c)$ similarly shows sizable reduction of both polarities with time.
Case $(d)$ shows reduction of the relatively strong negative-polarity
patch to the lower right of the \acp{QSEB}.

\paragraph{Visibility in other diagnostics.}
Figure~\ref{fig:diag} completes our presentation of the \acp{QSEB} phenomenon
by comparing their visibility in \Halpha\ with their visibility in
other spectral diagnostics, and also comparing \acp{QSEB} appearance with
facular striations (``pseudo-QSEBs'').

Cases $(a)$ and $(b)$ show \acp{QSEB}s without signature in the optical
continuum, nor in \FeIline\ and in AIA and IRIS ultraviolet imaging. 

Case $(c)$ suggests that there is a tiny continuum bright point
at the foot of the \acp{QSEB} flame with corresponding brightening of
\CaIR\ and \FeIline.
The \CaIR\ profile also shows higher wing intensities than in the
reference spectrum, but the increase remains below the minimum 140\%
wing brightening threshold defined for \acp{EB}s in \pII; the
brightening in the image cutout is indeed less than for other magnetic
concentrations nearby.
 Similarly, \CaIR\ wing for case $(b)$ shows a small bright point at the site of the \acp{QSEB} but not nearly at the high contrast as in the \Halpha\ wing. The \CaIR\ wings stay below the 120\% as compared to the reference (not shown for this case).
Detailed CRISPEX inspection of \CaIR\ behavior at the sites of other
\acp{QSEB}s confirmed that they do not show up as noticeable features in
this line.

Case $(d)$ serves to compare facular brightening with \acp{QSEB}
brightening. 
The facular striation morphology is identical between the \Halpha\
wing, the continuum, the core of \FeIline, and even in the ultraviolet
images at their much coarser resolution.  
One might expect that the small size of \acp{QSEB}s makes them
invisible at the 1.2~arcsec AIA resolution, but this example and more in
general the good visibility of \acp{MC}s in the AIA ultraviolet channels
suggests an intrinsic lack of ultraviolet continuum brightening for
\acp{QSEB}s.

The \Halpha\ profile indeed suggests continuation into continuum
brightening beyond the CRISP spectral sampling extent.

\section{Discussion}   \label{sec:discussion}

\paragraph{Small reconnection events.}
\acp{QSEB}s have similar bright-flame appearance in the \Halpha\ wings as
\acp{EB}s (Figs.~\ref{fig:overview}--\ref{fig:wbfast}), but they occur in
quiet areas instead of active regions (Figs.~\ref{fig:overview},
\ref{fig:wbfast}--\ref{fig:diag}), they are less bright and less tall
(Figs.~\ref{fig:overview}--\ref{fig:wbfast}), and they differ
significantly in their spectral visibilities (Fig.~\ref{fig:diag}).  
Nevertheless, they probably also mark small-scale magnetic
reconnection in the photosphere (Fig.~\ref{fig:stokes}).

It seems that there is an extended range of small reconnection events
taking place in the low solar atmosphere, with variation in magnetic
topography, amount of reconnection, released energy, and penetration
into the the higher atmosphere.
At the top end there are small flaring active-region fibrils (FAF),
very bright in AIA 1600\,\AA\ images, often producing ``IRIS bombs'',
and affecting the atmosphere above the chromospheric canopy
(\citeads{2009ApJ...701.1911P}; 
\citeads{2013JPhCS.440a2007R}; 
\citeads{2014Sci...346C.315P}; 
\pIII; \citeads{2016A&A...590A.124R}; 
\citeads{2016arXiv160405423T}). 

Next are \acp{EB}s which are also limited to complex bipolar active regions
but only rarely penetrate through the canopy (there are 
reports of accompanying surges but these are rare, see
\citeads{2002ApJ...575..506G}; 
\pI, \pII,
\citeads{2015ApJ...805...64R}). 
Nevertheless, also the sub-canopy heating in \acp{EB}s can be large enough
to cause significant brightening of the ultraviolet IRIS lines (\pIII,
\citeads{2016arXiv160405423T}). 

The \acp{QSEB}s reported here represent a yet weaker member of this family
occurring in quiet areas and without signature at ultraviolet wavelengths.

\paragraph{Spectral visibilities.}
\acp{QSEB}s share key spectral characteristics with \acp{EB}s: they show up as
intense wing brightenings of \Halpha\ without affecting the \Halpha\
core, they do not show up in the optical continuum, and they do not
show up in neutral-atom lines, at least not in \FeIline.
However, they differ distinctly from \acp{EB}s in other characteristics since
they do not show up in the \CaIR\ line nor in the AIA and IRIS
ultraviolet continua, whereas \acp{EB}s and also \acp{MC}s do show up in these
diagnostics.
In addition, \acp{EB}s can show strong emission in ultraviolet
``transition-region'' lines including even \ion{Si}{iv}
(\pIII; \citeads{2016arXiv160405423T}). 

Let us first discuss \acp{MC} visibilities.
The most detailed analysis of \acp{MC} ``line gap'' brightening of \FeI\
lines in top-down viewing is in
\citetads{2009A&A...499..301V}; 
it results from enhanced \FeI\ ionization in low-density fluxtubes.
\acp{MC} brightenings as ``G-band bright points'' result similarly from CH
dissociation (\citeads{2001ASPC..236..445R}, 
\citeads{2004ApJ...610L.137C}). 
\acp{MC} brightening of the \Halpha\ wings also results from relatively low
gas density but not by ionization but by less collisional damping
(Leenaarts et al., 
\citeyearads{2006A&A...452L..15L}; 
\citeyearads{2006A&A...449.1209L}). 

The good \acp{MC} visibility in ultraviolet continua results also from
ionization of \FeI\ and the other neutral ``electron donor'' species
since their bound-free edges dominate the ultraviolet continuous
extinction in the upper photosphere.
Hence, in top-down viewing the ultraviolet \acp{MC} radiation originates
much deeper than the height of formation suggested by standard
one-dimensional non-magnetic solar atmosphere models.
It is a misconception to attribute \acp{MC} ultraviolet brightness to
heating; it is also ``hole-in-the-surface'' radiation as in 
\citetads{1976SoPh...50..269S}. 
The observed morphology samples the low photosphere; indeed, near the
center of the disk the ultraviolet bright points correspond closely to
\acp{MC}s in photospheric magnetograms, a useful property for coaligning HMI
and AIA imagery also exploited here.

The same visibility mechanisms apply in slanted viewing towards the
limb where \acp{MC}s appear as faculae, again with \acp{MC} tenuity causing
fluxtube transparency in minority-stage lines and edges through
ionization, in the \Halpha\ wings through less damping. 
The cutout images on bottom row $(d)$ of Fig.~\ref{fig:diag} show this
very well: the morphology of the facular striae is remarkably similar
between these diagnostics.

\acp{EB}s also show up in ultraviolet continua but much brighter than \acp{MC}s
(\pII). 
The \acp{EB} visibility analysis by
\citetads{2016A&A...590A.124R} 
suggests that they are so hot (over 10,000\,K) that Balmer continuum
extinction, which increases very steeply with temperature, exceeds the
extinction of the metal edges and \Hmin.

The \acp{QSEB} visibilities represent an intermediate class.
The absence of visibility in \FeIline\ combined with their
low-photosphere anchoring (shown by flame feet being located in
intergranular lanes when viewed from aside, best seen in
Fig.~\ref{fig:wbfast}) suggests sufficient heating to ionize \FeI\ --
but in gas that is denser than the surroundings, not more tenuous as
is the case in \acp{MC}s.
Reconnection simulations as the one of
\citetads{2009A&A...508.1469A} 
indeed predict sizable density increase in such events.

The \acp{QSEB} invisibility in the ultraviolet continua may then be
explained by the gain in \Hmin\ extinction from the density increase
offsetting the loss of metal-edge extinction from ionization, but
without getting as hot as \acp{EB}s and therefore not passing into
Balmer-continuum domination.

The lack of significant wing brightening in \CaIR\ suggests that the
heating in \acp{QSEB}s does not reach the heights where this line forms,
whereas it does in \acp{EB}s.
It is therefore of interest to observe \acp{QSEB}s also in the
\ion{Na}{i}\,D and \ion{Mg}{i}\, b lines which form at intermediate
heights (\pIV,
\citeads{2016A&A...590A.124R}). 

\acp{EB} visibilities already define a complex set of constraints that is
hard to explain by classical static modeling and probably requires
time-dependent non-equilibrium modeling
(\citeads{2016A&A...590A.124R}). 
The same is likely the case for the \acp{QSEB} visibilities reported here.

\paragraph{Temperatures.}
The extinction curves in Figure~5 of
\citetads{2016A&A...590A.124R} 
together with the \acp{QSEB} and \acp{EB} visibilities suggest that both \acp{QSEB}s and
\acp{EB}s form at high density (total hydrogen density about
$10^{15}$~cm$^{-3}$) but at temperatures around 8,000\,K in \acp{QSEB}s,
causing visibility in the \Halpha\ wings but invisibility in the
Balmer continuum, and at temperatures above 10,000~K for \acp{EB}s causing
visibility also in the Balmer continuum and the IRIS lines.

\paragraph{Modeling.}
There have been \Halpha\ best-fit \acp{EB} modeling efforts based on
perturbing a standard model atmosphere in 1D or 2D geometry
(\citeads{1983SoPh...87..135K}; 
\citeads{2010MmSAI..81..646B}; 
\citeads{2013A&A...557A.102B}; 
\citeads{2014A&A...567A.110B}) 
that proposed \acp{EB} temperatures below 10,000~K.
These have been criticized for not being able to explain observed EB
brightness in the IRIS lines (\pIII; \pIV;
\citeads{2016A&A...590A.124R}; 
\citeads{2016arXiv160405423T}). 
We now suggest that these models may be more appropriate to describe
\acp{QSEB}s rather than \acp{EB}s.   
The same may hold for the \acp{EB} simulation efforts in
\citetads{2013ApJ...779..125N} 
and \citetads{2015ApJ...805...64R}. 

\paragraph{Energy release}
Since all our observed \acp{QSEB}s have \Halpha\ cores that are dominated by
overlying chromospheric fibrils we cannot establish what the intrinsic
\Halpha-core profiles of \acp{QSEB}s are, just as is the case for \acp{EB}s.
The \Halpha\ opacity of these features must increase very much towards
line center so that one may expect intrinsic profiles that are
comparable to those of other very strong lines, in particular
\ion{Mg}{ii}~k which has similar line-center extinction as \Halpha\ in
the \acp{EB} parameter range proposed by
\citetads{2016A&A...590A.124R}. 
We may therefore expect a profile underneath the canopy resembling
\ion{Mg}{ii}~k in shape: having high twin peaks with a central
self-absorption dip from resonance scattering.
The line core cannot be observed since blocked by the fibril canopy,
but underneath that it represents a bright radiator.

Estimation of \Halpha\ radiation loss based on observed extended-wing
emission excess and assuming optically thin line formation, as done
recently for \acp{EB}s by \citetads{2016arXiv160307100R}, 
ignores this intrinsic core contribution and that the
features are optically thick already in the observed \Halpha\ wing
enhancements since \acp{EB}s and \acp{QSEB}s appear intransparent.

In our opinion, quantitative estimation of the radiation losses in
these features first requires convincing duplication of their
characteristics and spectral visibilities in a numerical MHD
simulation, then detailed analysis of the radiation budget in such a
simulation. 
However, since \acp{QSEB}s and most \acp{EB}s do not affect the chromospheric
canopy nor the higher atmosphere, such evaluation seems unimportant
in the context of chromospheric or coronal heating in general. 

\paragraph{Reconnection scenario.}
The magnetogram sequences in Fig.~\ref{fig:stokes} suggest
reconnection as \acp{QSEB} cause by showing complex bipolar field patterns
with considerable evolutionary change.
However, the topography is less clear-cut then for \acp{EB}s which at the
best seeing in \acp{SST} magnetometry always mark cancelation of small
patches of field that move fast into larger patches of opposite
polarity (\pII), usually at sites with fast and sheared streaming
motions and correspondingly elongated granules. 
Even at the low AIA resolution fast movie playing of its 1700\,\AA\
sequences, which we have done for many emerging active regions, shows
invariably that \acp{EB}s occur at sites where \acp{MC}s move together, usually
with one (visible as a bright 1700\,\AA\ pixel) coming in fast from
far away.
For \acp{EB}s in emerging active regions these characteristics suggest
confirmation of the large-scale serpentine U-loop emergence scenario
advocated by, e.g.,
\citetads{2002SoPh..209..119B}, 
\citetads{2004ApJ...614.1099P}, 
\citetads{2007ApJ...657L..53I}, 
\citetads{2009A&A...508.1469A}, 
\citetads{2009ApJ...701.1911P}. 

For \acp{QSEB}s we may consider small-scale Omega-loop emergence with a
mass-collecting and reconnecting ``dipped'' sag between the anchored
opposite-polarity feet, as in cartoon models (a) in Fig.~12 of
\citetads{2002ApJ...575..506G} 
and 2a in Fig.~17 of \citetads{2008ApJ...684..736W}. 
The major difference with \acp{EB}s may be that in active-region
serpentine emergence the mass drainage occurs over much larger loop
lengths, feeding larger reconnection events.

\paragraph{Nomenclature.}
It has been suggested to us that we should not increase the taxonomic
categories of solar features with yet another one.  
However, we feel that simply calling \acp{QSEB}s ``EBs in quiet
areas'' would be dangerously misleading. 
The \acp{EB} literature contains many publications that mistook
quiescent \acp{MC}s for \acp{EB}s. 
The risk that such misinterpretation affect \acp{QSEB}s is even larger
because they are intrinsically weaker than \acp{EB}s. 
Their small-flame morphology may currently be rendered distinctively
only by the \acp{SST}.
Also, Ellerman's original description defines \acp{EB}s as brilliant
\Halpha-wing brightenings in complex active regions only.
Furthermore, \acp{QSEB}s differ intrinsically from \acp{EB}s by not
appearing as beyond-\acp{MC} brightening in \CaIR\ nor showing up in
ultraviolet continua.
We therefore maintain our naming as \acp{QSEB}, a new solar
phenomenon.

\section{Conclusion}   \label{sec:conclusion}
We have described a new quiet-Sun phenomenon. 
It was not reported so far because it is very elusive: \acp{QSEB}s can only
be distinguished from near-limb facular striations in imaging
spectroscopy that reaches the unsurpassed quality of the \acp{SST} at the
best La Palma seeing.

\acp{QSEB}s probably originate from magnetic reconnection in the low
photosphere and therefore promise to be valuable diagnostics to chart
quiet-Sun field topography evolution.

The spectral visibilities of the \acp{QSEB}s are complex. 
In some respects they are similar to those of active-region \acp{EB}s
(extended wing brightenings of \Halpha, no brightening of the optical
continuum, no brightening of \FeI\ lines), but they differ
significantly in other respects (no striking brightening of the
wings of \CaIR, no brightening of the AIA 1600 and 1700\,\AA\
continua, no brightening in the IRIS ultraviolet slitjaw images).

The magnetic topography of the quiet areas where they occur suggests a
reconnection scenario which differs from the one causing \acp{EB}s in
emerging active regions.

As is the case for \acp{EB}s, advanced time-dependent radiation-MHD
simulations seem required to understand the details of \acp{QSEB} formation. 
Similarly to \acp{EB}s
(\citeads{2016A&A...590A.124R}), 
the synthesis of the various spectral diagnostics needed to reproduce
and explain the diverse visibilities must probably account for
non-equilibrium hydrogen ionization and recombination. 
The reward of such endeavor will be that the complex \acp{QSEB} occurrence
and visibility patterns will then provide sufficient constraints for
definitive identification and analysis of the underlying reconnection
processes. 

\begin{acknowledgements}
The Swedish 1-m Solar Telescope is operated on the island of La Palma
by the Institute for Solar Physics of Stockholm University in the
Spanish Observatorio del Roque de los Muchachos of the Instituto de
Astrof{\'\i}sica de Canarias.
Our research has been partially funded by the Norwegian Research
Council and by the ERC under the European Union's Seventh Framework
Programme (FP7/2007-2013)\,/\,ERC grant agreement nr.~291058.
This work was initiated after discussions at the meeting ``Solar UV
bursts - a new insight to magnetic reconnection" at the International
Space Science Institute (ISSI) in Bern.
IRIS is a NASA small explorer mission developed and operated by LMSAL with mission operations executed at NASA Ames Research center and major contributions to downlink communications funded by ESA and the Norwegian Space Centre.
%
We thank A. Drews, S. Jafarzadeh, T. Leifsen, T. Pereira, 
A. Sainz-Dalda, M. Szydlarski, R. Timmons, and P. Zacharias for their
help with acquiring the observations.
T. Pereira performed the alignment of the 2014 June 17 IRIS and AIA
observations to the \acp{SST} data. 
We made much use of \acp{NASA}'s Astrophysics Data System Bibliographic
Services.

\end{acknowledgements}



\end{document}